\documentclass[prl,showpacs,twocolumn,superscriptaddress]{revtex4-1}
\usepackage{amsfonts}
\usepackage{amsmath}
\usepackage{amssymb}
\usepackage{graphicx}
\usepackage{epsfig}
\usepackage{bm}

\begin{document}


\title{Experimental characterization of qutrits using SIC-POVMs}




\author{Z. E. D. Medendorp$^{1,\dagger}$, F. A. Torres-Ruiz$^{2,3}$, L. K. Shalm$^1$, G. N. M. Tabia$^4$, C. A. Fuchs$^4$, and A. M. Steinberg$^1$
\medskip
\\
\small
$^1$Centre for Quantum Information $\&$ Quantum Control and Institute for Optical Sciences, Department of Physics, University of Toronto, 60 St George Street, Toronto, Ontario, Canada M5S 1A7
\\
\small
$^2$Center for Optics and Photonics, Universidad de Concepci\'on, Casilla 160-C, Concepci\'on, Chile
\\
\small
$^3$Departamento de Ciencias F\'{i}sicas, Universidad de La Frontera, Temuco, Casilla 54-D, Chile.
\\
\small
$^4$Perimeter Institute for Theoretical Physics, Waterloo, Ontario N2L 2Y5, Canada
\\
\small
$^\dagger$zmedendo@lakeheadu.ca}

\begin{abstract}
Generalized quantum measurements (also known as POVMs) are of great importance in quantum information and quantum foundations, but often difficult to perform.  We present an experimental approach which can in principle be used to perform arbitrary POVMs in a linear-optical context.  One of the most interesting POVMs, the SIC-POVM, is the most compact, set of measurements that can be used to fully describe a quantum state.  We use our technique to carry out the first experimental characterization of the state of a qutrit using SIC-POVMs.  Because of the highly symmetric nature of this measurement, such a representation has the unique property that it permits all other measurement outcomes to be predicted by a simple extension of the classical Bayesian sum rule, making no use of complex amplitudes or Hilbert-space operators.  We demonstrate this approach on several qutrit states encoded in single photons.

\end{abstract}


\maketitle

In recent years, it has become widely appreciated that the projective valued measurements (PVM) described in quantum mechanics textbooks are merely a restricted class of the physically possible measurements, known as ``generalized measurements'' or positive operator-valued measures (POVMs).  For certain tasks, non-projective POVMs have been shown to be superior; for instance, they achieve the theoretical bound for unambiguous state discrimination \cite{IDP,barnett}, and are potentially of vital importance in quantum cryptographic applications \cite{Crypto1}.  It is not straightforward to directly implement a desired POVM in the lab, however.  In principle, one can always do so by performing a suitable coupling between the system and an ancillary system with a large number of degrees of freedom, and then projectively measuring the latter apparatus \cite{POVM2}; in practice, the appropriate coupling is often unavailable.  One can often ``mock up'' a POVM by cycling through a number of different projective measurements, but this can often be inefficient, in a way we will describe below.  Motivated by the recent observation that symmetric, informationally complete POVMs (SIC-POVMs) provide a uniquely elegant description of quantum states \cite{Fuchs1}, and the most efficient technique for characterizing quantum states in all dimensions \cite{Scott}, we have developed a method which allows arbitrary POVMs to be approximated arbitrarily well in linear optics, using only presently available technology.  We demonstrate this technique by performing the first SIC-POVM characterization of optical qutrits, and illustrate the novel features of this resulting representation.

\begin{figure}[ht!]
\centering
\epsfig{file=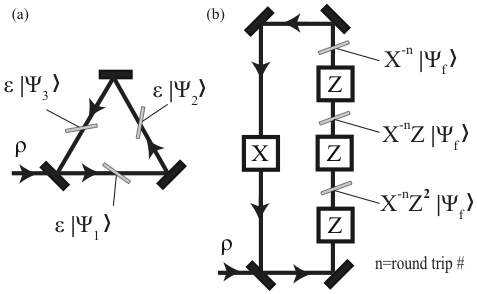,scale=0.8}
\caption{\setlength{\baselineskip}{8pt}{(a) POVM with 3 elements inside a storage loop (b) Compacted version of the POVM scheme to implement a 9-element SIC-POVM for $d=3$ over three round trips. The labels refer to the interpretation of a click.}}
\label{fig:Fig1}
\end{figure}

\begin{figure}[hb!]
\centering
\epsfig{file=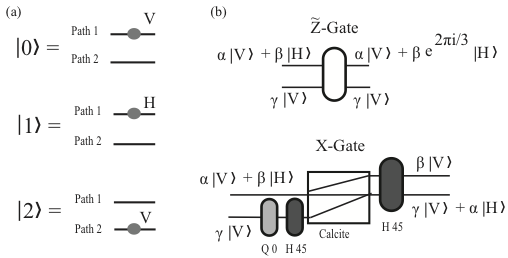,scale=0.9}
\caption{\setlength{\baselineskip}{8pt}{(a) Qutrit encoding and our fiducial state for $d=3$ (b) The \~Z gate applies a $\frac{\lambda}{3}$ phase shift between H and V polarization. The X gate cycles through the three basis states.}}
\label{fig:IdealFig1}
\end{figure}

\begin{figure*}[ht!]
\centering
\epsfig{file=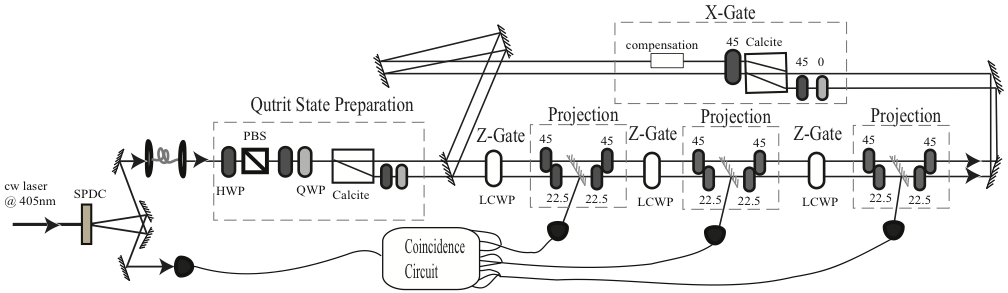,scale=0.9}
\caption{\setlength{\baselineskip}{8pt}{SIC-POVM schematic where SPDC generates a signal photon to be encoded as a qutrit, and an idler photon provides timing information to determine which of the 3 possible round trips the signal photon exited the spiral.  HWPs ensure the correct rotations for state measurements. }}
\label{fig:FigApp}
\end{figure*}

The state of an ensemble of $d$-dimensional quantum systems is most typically described by a density matrix, an abstract list of $d^2$ numbers.  Quantum state tomography involves carrying out at least $d^2$ linearly independent measurements and then using one of a number of mathematical techniques to estimate the state.  Over the past several years, there has been a great deal of interest in determining the most efficient and robust sets of measurements for this task \cite{Kwiat1,Kurtsiefer,MUBs,QST1}.  Arguments suggest that the more symmetric the measurement set, the less biased the estimation will be \cite{LangAlt}.  In addition, the {\it smaller\/} the measurement set, the less redundancy there is, and the faster tomography will converge.  For instance, the standard set of measurements for a qubit can be represented as the $\pm x$, $\pm y$, and $\pm z$ projections on the Bloch sphere -- the so-called mutually unbiased bases (MUBs), as no one projection gives any information about the two other projections. Only recently has this been extended to more than one qubit \cite{MUBs}, because of the difficulty of generalizing these measurements to higher dimensions.  However, this characterization requires 6 different projections, while in principle 4 projections (the vertices of a regular tetrahedron embedded in the Bloch sphere) would be the minimal sufficient set.  This tetrahedral set of projectors is the SIC-POVM for a qubit \cite{Kurtsiefer,SICPOVM2}.  One experiment has recently used SICs to perform qubit tomography \cite{Kurtsiefer}, but extensions to higher-dimensional systems have proved challenging.  Is it possible in practice to measure a higher-dimensional POVM without necessarily including unwanted extra projectors?

This prospect has more than merely technical relevance.  A number of workers \cite{Hardy} have underscored that instead of an abstract list of complex numbers such as a density matrix, a list of $d^2$ directly measurable probabilities may serve as a complete characterization of a quantum state.  While any linearly independent set would suffice in principle, one of us \cite{Fuchs1} has recently observed that the outcomes of SIC-POVMs form a more natural description of the quantum state. If an observer uses SIC-POVMs to build up his statistics, then the Born rule takes on an elegant form as a quantum variant of the classic Bayesian sum rule,
\begin{equation}\label{eqn:QLTP}
p(B_j)=(d+1)\sum_{i=1}^{d^2} p(A_i) p(B_j|A_i) - 1.
\end{equation}
Here  $p(A_i)$ are the SIC-POVM probabilities, $p(B_j|A_i)$ are the probabilities for getting outcomes $B_j$ in a further von Neumann measurement if a SIC outcome $A_i$ were obtained first, and $p(B_j)$ are the probabilities for $B_j$ in a direct measurement of the observable. We call this form of the Born rule, the Quantum Law of Total Probability (QLTP)---to test it is to test the Born rule itself.  Moreover, the consistency of the QLTP is enough to imply much of the structure of quantum state space, where the $p(A_i)$ are restricted to a nontrivial convex set of all probability distributions \cite{Appleby}.

One could measure one POVM element by coupling it out with a polarizing beam splitter (PBS), but then only the orthogonal component of the state would remain, precluding the measurement of later elements.  If we instead use a low-reflectivity ``partially polarizing beam-splitter'' (PPBS) then a partial projection onto each element is possible, leaving the remaining state mostly unaltered. To implement a SIC-POVM, consider a sequence of $d^2$ partial projectors, where the remaining state from each SIC-POVM element is recycled into the next element.  If we place the sequence of elements into a lossless storage loop with an optical switch on the first mirror to accommodate photons entering the cavity, we can further recycle the state.  In fact this scheme can be used to perform any optical POVM.  Figure \ref{fig:Fig1}(a) shows a storage loop containing three partial projectors with a reflection of $\epsilon$. In principle, by sending the quantum state $\rho$ into the storage loop and by quickly operating the switchable input mirror we trap the photon inside, and it can only escape through the weakly coupled ports.  Since the projections are weak, the quantum state is nearly unaltered as it propagates through the loop.  In the limit that $\epsilon$ goes to zero the light will eventually leak out of one of the ports, and the time-integral of each port is taken to stand for a single POVM projector.  By replacing the three POVM elements with $d^2$ SIC elements a SIC-POVM is realized.

In our work, a SIC-POVM with elements $|\psi_{mn}\rangle$, $m,n\in\{0,1,2,\ldots,d-1\}$, is generated from a carefully chosen fiducial state $|\psi_f\rangle$ by actions of the shift and phase operators, $X$ and $Z$, via $|\psi_{mn}\rangle = X^m Z^n |\psi_f\rangle$
where $Z\left| j\right\rangle=e^{2 \pi i j /d}\left| j\right\rangle$,  $X\left| j\right\rangle=\left| j\oplus 1\right\rangle$ and $d$ is the dimensionality of the Hilbert space \cite{SICPOVM3}. For a qutrit ($d=3$), we use $|\psi_f\rangle=\frac{1}{\sqrt{2}}(|0\rangle+|1\rangle)$.

A compact measurement device is constructed with $d+1$ gates, three Z and one X ($d=3$), which couple light for all nine SIC elements in $d$ round trips, see Figure \ref{fig:Fig1}(b).  The gates permute the measurement basis from each of the nine elements to the fiducial state, and are measured with polarization-dependent reflections.  For example, measuring the $|\psi_{20}\rangle$ element requires the action of $X$ to rotate to $|\psi_{00}\rangle=|\psi_{f}\rangle$.  A photon exiting after one round trip and after the third $Z$ gate represents such a measurement (ie. $Z^3XZ^3=X$).

Our signal photon is generated in spontaneous parametric down conversion (SPDC), and the idler photon is used for timing information.  As shown in FIG. \ref{fig:IdealFig1}(a) the qutrit is encoded in the path and polarization of the signal: $|0\rangle = |V_{1}\rangle$, $|1\rangle = |H_{1}\rangle$ and $|2\rangle = |V_{2}\rangle$; the state $|H_{2}\rangle$ is not used.  Using wave plates and a calcite beam displacer (BD) we can prepare arbitrary pure states of the qutrit. For convenience, where only passive components are required, the qutrit now passes over the top of the first mirror and enters a spiral loop where it  can traverse the gates and measurements several times.  In our proof-of-principle experiment only one pass of the nine projectors is realized, requiring three round trips of the loop shown in Fig. \ref{fig:FigApp}.

The experimental implementations of the gates are shown in Fig. \ref{fig:IdealFig1}(b).  A liquid crystal wave plate (LCWP) applies a $\frac{2\pi}{3}$ phase shift between horizontal and vertically polarized light to perform the \~Z operation. This differs from the ideal $Z$ operation in that the second path does not experience any phase shift (\~Z$ = \exp(-4\pi i \delta_{j2}/3)Z$). Since the second path is never directly measured (there is no $j=2$ content in the fiducial state), \~Z$^n|\psi_f\rangle = Z^n|\psi_f\rangle$, and as long as \~Z$^3=I$ is satisfied, the loop functions properly.  Three half wave plates (HWPs) and a BD facilitate the cycling through of the basis states yielding the $X$ operation. Note that the paths have switched places after exiting the $X$ gate and are returned to the correct orientation before continuing through the spiral.

\begin{figure*}[ht!]
\centering
\epsfig{file=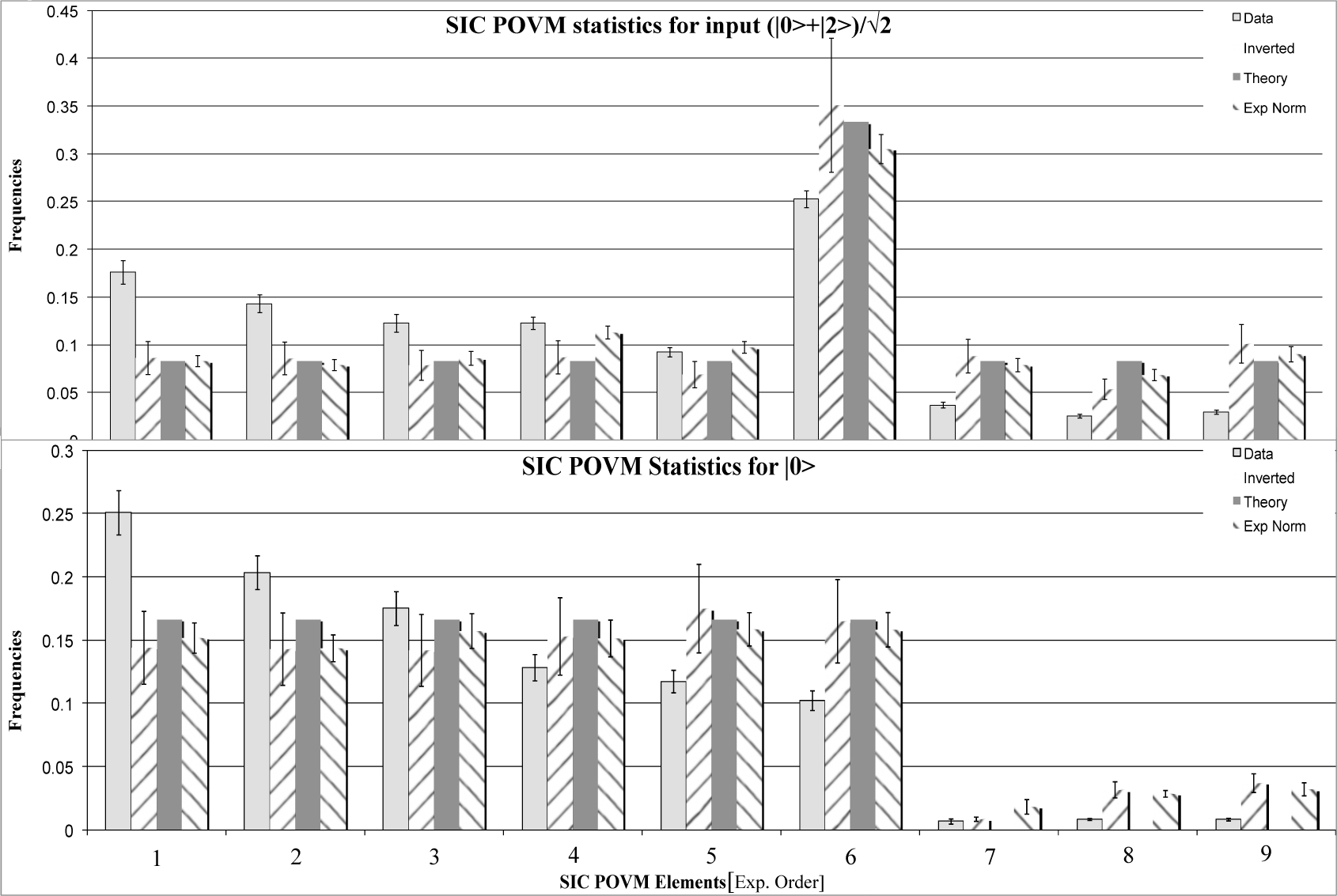,scale=0.2}
\caption{\setlength{\baselineskip}{8pt}{Comparison of SIC element frequencies for input a)$\frac{1}{\sqrt{2}}\left(\left|0\right\rangle+\left|2\right\rangle\right)$ b) $|0\rangle$, between raw data, theory and two correction methods. The SIC-POVM elements are listed in the order experienced by the photon. Statistical noise and systematic errors are included in the error bars.}}
\label{fig:Fig2}
\end{figure*}

Measurement of the state is performed by a collection of HWPs and a PPBS (a glass slide near Brewster's angle) after each $Z$ gate. Since measurements are realized on the first path alone, coherence between the paths is significant only for the $X$ gate.  By balancing the components in each path and matching the BDs in the qutrit generation with the $X$ gate, coherence is automatically ensured for the first two round trips of the loop.  The third round trip is balanced using custom thickness pieces of BK7.  The visibilities after the second and third round trips dropped to 74\% and 36\%, respectively, due to the outcoupling and other losses in the loop.

The outcomes for two different inputs are chosen to display the characteristics of our SIC-POVM apparatus. Figures \ref{fig:Fig2} a) and b) both reveal an exponential decay associated with losses in the loop.  Since input state $\frac{1}{\sqrt{2}}\left(\left|0\right\rangle+\left|2\right\rangle\right)$ is one of the SIC elements, the overlap is expected to be $\frac{1}{3}$ with the $6^{th}$ element and $\frac{1}{12}$ with all other elements.   For input state $|0\rangle$ the first six outputs should have equal probability of $\frac{1}{6}$, and the last three should be zero since this input is orthogonal to these three SIC elements. Deviations from zero are due to poor visibility at the X gate for the final round trip, which is caused by the lack of collimation of the beam after traversing this far. Because of decay, the raw data does not have the expected shape, and therefore must be corrected.

Our PPBS reflected $13\%$ of the vertical and $3\%$ of the horizontal polarization, leading to significant deviation from the ideal low-coupling limit.  We compared two methods for correcting for such effects.  In one, we fully modelled the experiment to determine the best-fit density matrix for a given data set.  This density matrix allows us to recover the SIC-POVM elements, plotted as the Inverted data in Figure \ref{fig:Fig2}.  In another, we use only an experimentally determined normalization correction: we scaled each POVM output to the count rate observed for a completely mixed input state, to account for the state-dependent loss.  The resulting POVM elements are plotted as the Exp. Norm data in Figure \ref{fig:Fig2}.  Although both methods appear to account for the imperfections adequately, the experimental correction has smaller statistical uncertainties.

Systematic errors are estimated by a Monte-Carlo simulation of a one degree misalignment of the HWPs.  The figures show that there is good agreement between theory and experiment for the first 4 or 5 measurements.  The remaining measurements deviate more significantly because the beam does not remain collimated for the full 5m of the 3 pass loop and this affects both the fiber coupling and the X Gate efficiency.

To test the QLTP, Eq.~(\ref{eqn:QLTP}) requires the SIC-POVM outcomes, PVM outcomes for a particular projection of all nine SIC elements, and for comparison, the PVM outcome of the input state directly.  (See \cite{Born1} for another test of the Born rule.)  In addition to the SIC-POVM, a tomography apparatus was built beside the qutrit generation so that the beams could be diverted from the SIC-POVM to the tomography setup.  The first set of measurements are the PVM outcomes for each SIC element.  This information is gathered by preparing each of the SIC in the qutrit generation and performing PVMs, for example projecting onto $\frac{1}{\sqrt{2}}(|0\rangle +i|2\rangle)$.  Data acquisition time was $50s$, coincidence window was set to $4ns$, and the accumulation of coincidences in both orthogonal ports summed to $10^5$. This measurement was done nine times, once for each SIC element.  Next, an input state was prepared and tomography was performed on it with the same settings.
   The PVM of interest was also performed.  Finally, we send the state into the SIC-POVM and collect for $200s$ to accumulate $3$ to $6\times10^4$ coincidences, depending on the input.

Using the experimentally corrected SIC-POVM outcomes, we tested the QLTP for several input states.  For example, for
the input $\frac{1}{\sqrt{2}}(|0\rangle +|2\rangle)$, the QLTP predicted the probability of measuring $\frac{1}{\sqrt{2}}(|0\rangle +i|2\rangle)$ to be  0.526 $\pm$ 0.009(stat) $\pm$ 0.03(sys), while in a direct measurement we found 0.506 $\pm$ 0.004 $\pm$ 0.002.  For the input $|0\rangle$, the QLTP predicted the probability of measuring $\frac{1}{\sqrt{2}}(|1\rangle +|2\rangle)$ as 0.01 $\pm$ 0.0002(stat) $\pm$ 0.009(sys), while in a direct measurement we found 0.005 $\pm$ 0.0004 $\pm$ 0.001.

The technique we presented here can be used to measure any desired linear-optical POVM, and existing technology would permit such measurements to be done with high accuracy.  We have shown that the ability to characterize quantum systems using SIC-POVMs enables tomography to be performed in the most mathematically compact manner possible, and that representation in terms of SIC-POVM elements enables the Born rule to be recast as a simple generalization of the classic Bayesian sum rule.  This offers both a new way to measure quantum states and a new way to think about them.

\begin{acknowledgments}
We thank NSERC, CIFAR, FATR, CAF, QuantumWorks, U.~S. ONR Grant No.\ N00014-09-1-0247, and PBCT-Red 21 for support.  Additional thanks go to Alan Stummer for designing the coincidence circuit, and Lee Rozema for assistance with the figures.
\end{acknowledgments}

\end{document}